\documentstyle[amssymb,aps, manuscript]{revtex}
\begin{document}
\title{The Absence Of Saturation Of The Level Number Variance In A Rectangular Box}
\author{J. M. A. S. P. Wickramasinghe and R. A. Serota}
\address{Department of Physics\\
University of Cincinnati\\
Cincinnati, OH\ 45221-0011}
\maketitle

\begin{abstract}
The variance of the number of levels in an energy interval around a level
with large quantum numbers (semiclassical quantization) is studied for a
particle in a rectangular box. Sampling involves changing the ratio of the
rectangle's sides while keeping the area constant. For sufficiently narrow
intervals, one finds the usual linear growth with the width of the interval.
For wider intervals, the variance undergoes large, non-decaying oscillations
around what is expected to be the saturation value. These oscillations can
be explained as a superposition of just a few harmonics that correspond to
the shortest periodic orbits in the rectangle. The analytical and numerical
results are in excellent agreement.
\end{abstract}

\section{Introduction}

Two decades ago, Casati, Chirikov and Guarneri\cite{CCG} and Berry\cite{B}
have undertaken numerical and analytical studies of the level statistics in
rectangular billiards for large quantum numbers (semiclassical
quantization). Casati {\it et al}\cite{CCG} demonstrated numerically that
the distribution function of the nearest-neighbor energy level spacings is
exponential (obeys Poisson law\cite{BFFMPW}), which is generally the case
for classically integrable systems and is well understood theoretically\cite
{G}. Further, they studied the behavior of the level rigidity\cite{BFFMPW}
in the spectral staircase and showed that, following the initial linear
growth with the width of the interval, the rigidity saturates to a constant
value. This behavior was explained by Berry\cite{B} who argued that the
width of the interval at which the saturation occurs corresponds to the
shortest periodic orbit in the billiard and that the saturation value of the
rigidity can be obtained as a sum over the periodic orbits. Subsequently,
the results of\cite{CCG} have been reproduced with high numerical precision
in Refs.\cite{Gr} and\cite{RV}.

In the previous work\cite{SW}, we set out to interpret this result in terms
of the global level rigidity and, in particular, proposed an ansatz for the
level density correlation function that successfully describes the
transition from linear to saturation behavior and that is also similar,
aside from the energy scale, to its counterparts in the Gaussian ensembles
(classically chaotic systems)\cite{BFFMPW},\cite{G}. Furthermore, our
correlation function led us to believe that the variance should exhibit
interesting, albeit decaying, oscillations on approach to its saturation
value (by the integral relationship\cite{BFFMPW} between the rigidity and
the variance - see below - it should be twice that of the rigidity).

Surprisingly, our numerical calculation indicated that the size of the
oscillations of the variance does not decay with the increase of the width
of the interval. Further investigation of the level correlation function
revealed that a more accurate treatment of large energy scales gives an
analytical explanation which is in excellent agreement with the numerical
results.

It should be emphasized that the key difference of our numerical procedure
relative to\cite{Gr} and\cite{RV} is that we averaged over the ratio of the
rectangle's sides, as opposed to averaging over the spectrum keeping a fixed
ratio of the sides. Indeed, both the onset of saturation and the saturation
value scale as a square root of energy, that is depend critically on the
position in the spectrum\cite{B}. Consequently, averaging over the spectrum
averages over these quantities. Notice, for instance, that the plot of
variance in\cite{Gr} does show oscillations. However, because of the
averaging procedure, they appear to be decaying and of irregular pattern and
did not receive due acknowledgment.\footnote{%
We learned about Ref.\cite{Gr} after completion of this work.}

\section{Notations and Known Results}

We will consider intervals $\left[ \varepsilon -E/2,\varepsilon +E/2\right] $
, $E\ll \varepsilon $, where the states with energies near $\varepsilon $
have large quantum numbers and can be described semiclassically. We will
denote ${\cal N}\left( \varepsilon \right) $ the spectral staircase\cite{G} 
\begin{equation}
{\cal N}\left( \varepsilon \right) =\sum_{k}\theta \left( \varepsilon
-\varepsilon _{k}\right)  \label{cN}
\end{equation}
where $\theta $ is unit step function and $k$ are energy eigenstates. More
precisely, we will consider the 2D-reduced\cite{G} spectrum where the
systematic energy dependence of ${\cal N}\left( \varepsilon \right) $ has
been eliminated\footnote{%
For a finite domain, in particular, we must eliminate the systematic energy
dependence of $\left\langle {\cal N}\left( \varepsilon \right) \right\rangle 
$ due to boundary corrections \cite{G},\cite{BB},\cite{BH}; for a rectangle
this dependence can be obtained directly\cite{BH},\cite{WSG} from the energy
levels of a particle in a box, eq. (\ref{enm}) below, and is given by $%
\left\langle {\cal N}\left( \varepsilon \right) \right\rangle =\varepsilon
-2\beta \sqrt{\varepsilon /\pi }+1/4$, where $\beta =\left( \alpha
^{1/4}+\alpha ^{-1/4}\right) /2$, $\alpha $ is defined in eq. (\ref{alpha})
below and $\varepsilon $ is measured in units of$\ \Delta =2\pi \hbar
^{2}/mA $, eq. (\ref{Delta_vs_rhoav}) below.}. Towards this end, we make the
substitution 
\begin{equation}
\varepsilon \rightarrow \left\langle {\cal N}\left( \varepsilon \right)
\right\rangle  \label{epsilon_reduced}
\end{equation}
and, in particular, $\varepsilon _{k}\rightarrow \left\langle {\cal N}\left(
\varepsilon _{k}\right) \right\rangle $, where $\left\langle {\cal N}\left(
\varepsilon \right) \right\rangle $ is the ensemble average of ${\cal N}%
\left( \varepsilon \right) $ (in the limit of an infinitely large ensemble).
With such definition of energy ({\it reduced} spectrum), we have 
\begin{equation}
\left\langle {\cal N}\left( \varepsilon \right) \right\rangle \rightarrow
\left\langle {\cal N}\left( \varepsilon \right) \right\rangle =\varepsilon
=\left\langle \rho \left( \varepsilon \right) \right\rangle \varepsilon
\label{cNav_reduced}
\end{equation}
that is $\left\langle \rho \left( \varepsilon \right) \right\rangle =1$,
where 
\begin{equation}
\rho \left( \varepsilon \right) =\frac{d{\cal N}\left( \varepsilon \right) }{%
d\varepsilon }=\sum_{k}\delta \left( \varepsilon -\varepsilon _{k}\right)
\label{rho}
\end{equation}
is the density of states. In dimensional units, (\ref{cNav_reduced}) can be
written as

\begin{equation}
\left\langle {\cal N}\left( \varepsilon \right) \right\rangle =\left\langle
\rho \left( \varepsilon \right) \right\rangle \varepsilon  \label{cNav}
\end{equation}
where 
\begin{equation}
\left\langle \rho \left( \varepsilon \right) \right\rangle =\frac{mA}{2\pi
\hbar ^{2}}  \label{rhoav}
\end{equation}
$m/2\pi \hbar ^{2}\ $being the 2D density of states in the thermodynamic
limit and $A$ the system area.

For a rectangle, the energy eigenvalues are 
\begin{equation}
\varepsilon _{nm}=\frac{\pi ^{2}\hbar ^{2}}{2m}\left( \frac{n^{2}}{a^{2}}+%
\frac{m^{2}}{b^{2}}\right)  \label{enm}
\end{equation}
and averaging is performed over the ''$\alpha $-ensemble,'' that is by
sampling the values of 
\begin{equation}
\alpha =\frac{a^{2}}{b^{2}}  \label{alpha}
\end{equation}
where $a$ and $b$ are the rectangle's sides, while keeping $A=ab$, and
consequently $\left\langle \rho \left( \varepsilon \right) \right\rangle $,
constant. Numerically, we use algebraic numbers for $\alpha $ to avoid
accidental level degeneracies (we don't use transcendental numbers because
they are too close to rational\cite{G}).

In this work we will not discuss the distribution of the nearest-neighbor
energy level spacings, except to state that our computations\cite{WSG} are
congruent with those of\cite{CCG},\cite{Gr} and\cite{RV}. Here, we will
study the correlations and fluctuations in the energy spectrum and, towards
this end, the following measures will be used. First, it is the ''level
rigidity'' $\Delta _{3}$, defined\cite{B},\cite{BFFMPW} as the best linear
fit of the spectral staircase on the interval $\left[ \varepsilon
-E/2,\varepsilon +E/2\right] $ 
\begin{equation}
\Delta _{3}\left( \varepsilon ,E\right) =\left\langle 
%TCIMACRO{\QATOP{\min }{\left( A,B\right) } }
%BeginExpansion
{\min  \atop \left( A,B\right) }%
%EndExpansion
\frac{1}{E}\int_{\varepsilon -E/2}^{\varepsilon +E/2}d\varepsilon \left[ 
{\cal N}\left( \varepsilon \right) -A-B\varepsilon \right] ^{2}\right\rangle
\label{Delta3}
\end{equation}
Fig. 1 shows an example of such a fit.

For the number of levels $N$ on the interval $\left[ \varepsilon
-E/2,\varepsilon +E/2\right] $ 
\begin{equation}
N\left( \varepsilon ,E\right) ={\cal N}\left( \varepsilon +\frac{E}{2}%
\right) -{\cal N}\left( \varepsilon -\frac{E}{2}\right)  \label{N}
\end{equation}
the variance 
\begin{equation}
\Sigma \left( \varepsilon ,E\right) =\left\langle \left( N-\left\langle
N\right\rangle \right) ^{2}\right\rangle  \label{Sigma}
\end{equation}
is another measure of the fluctuations. Clearly, 
\begin{equation}
\left\langle N\right\rangle =\left\langle \rho \right\rangle E  \label{Nav}
\end{equation}
for an infinitely large ensemble.

Introducing the mean level spacing $\left\langle \Delta \right\rangle $ as 
\begin{equation}
\left\langle \Delta \right\rangle =E\left\langle \frac{1}{N}\right\rangle
\label{Delta}
\end{equation}
we find that, due to the $N$-fluctuations $\Sigma \propto E$ over the
intervals $E$ narrower than a certain energy scale, the following
relationship holds\cite{WSG} 
\begin{equation}
\left\langle \rho \right\rangle \left\langle \Delta \right\rangle =1+\frac{1%
}{\left\langle N\right\rangle }+O\left( \frac{1}{\left\langle N\right\rangle
^{2}}\right)  \label{rhoav_vs_Delta}
\end{equation}
For classically integrable billiards the latter holds for $E\ll \sqrt{%
\varepsilon \left\langle \rho \right\rangle ^{-1}}$ (see below), so that we
can neglect the difference between $\left\langle \Delta \right\rangle $ and $%
\left\langle \rho \right\rangle ^{-1}$ only for sufficiently large intervals 
$\left\langle N\right\rangle \gg 1$, in which case we can introduce the
limiting value of mean level spacing (''global mean'') as 
\begin{equation}
\Delta =\frac{1}{\left\langle \rho \right\rangle }=\frac{2\pi \hbar ^{2}}{mA}
\label{Delta_vs_rhoav}
\end{equation}
In what follows we will, unless mentioned otherwise, measure energies in
units of $\Delta $ (wherein $\Delta =\left\langle \rho \right\rangle ^{-1}=1$%
, as in eq. (\ref{cNav_reduced})).

General relationships can be established between the fluctuations measures
and the correlation function of the level density\cite{BFFMPW}, 
\begin{eqnarray}
K\left( \varepsilon _{1},\varepsilon _{2}\right) &=&\left\langle \delta \rho
\left( \varepsilon _{1}\right) \delta \rho \left( \varepsilon _{2}\right)
\right\rangle  \label{rho_corr} \\
\delta \rho \left( \varepsilon \right) &=&\rho \left( \varepsilon \right)
-\left\langle \rho \left( \varepsilon \right) \right\rangle  \label{deltarho}
\end{eqnarray}
regardless of the specific form of $K\left( \varepsilon _{1},\varepsilon
_{2}\right) $, for instance, 
\begin{equation}
\Sigma \left( \varepsilon ,E\right) =\int_{\varepsilon -E/2}^{\varepsilon
+E/2}\int_{\varepsilon -E/2}^{\varepsilon +E/2}K\left( \varepsilon
_{1},\varepsilon _{2}\right) d\varepsilon _{1}d\varepsilon _{2}
\label{Sigma_integral}
\end{equation}
Using these relationships one can further show that $\Sigma $ supersedes $%
\Delta _{3}$ via an integral relationship\cite{BFFMPW} 
\begin{equation}
\Delta _{3}\left( \varepsilon ,E\right) =\frac{2}{E^{4}}\int dx\left(
E^{3}-2xE^{2}+x^{2}\right) \Sigma \left( \varepsilon ,x\right)
\label{Delta3_vs_Sigma}
\end{equation}
which, again, does not depend on the particulars of the level statistics and
level correlations in the semiclassical energy spectrum.

For Gaussian ensembles, corresponding to classically chaotic ergodic
systems, the functional form of the level correlation function $K\left(
\varepsilon _{1},\varepsilon _{2}\right) $ is well understood\cite{BFFMPW}.
Denoting 
\begin{equation}
\varepsilon =\frac{\varepsilon _{1}+\varepsilon _{2}}{2}\text{, }\omega
=\varepsilon _{2}-\varepsilon _{1}  \label{epsilon_omega}
\end{equation}
we find that it can be written, in most general terms, as 
\begin{equation}
K\left( \varepsilon _{1},\varepsilon _{2}\right) =K\left( \omega \right)
=\Delta ^{-2}\left[ \delta \left( \frac{\omega }{\Delta }\right) -{\cal K}%
\left( \frac{\omega }{\Delta }\right) \right]  \label{K_Gauss}
\end{equation}
where 
\begin{equation}
\int_{-\infty }^{\infty }{\cal K}\left( x\right) dx=1  \label{Int_cK}
\end{equation}
and 
\begin{equation}
\int_{-\infty }^{\infty }K\left( \omega \right) d\omega =0  \label{Int_K}
\end{equation}
The interpretation of these formulas are as follows. The $\delta $-function
term in (\ref{K_Gauss}) describes uncorrelated levels. The ${\cal K}$%
-function describes ''level repulsion'' and has a scale of $\Delta $. The
integrals (\ref{Int_cK}) and (\ref{Int_K}) converge to their values on the
scale of $\omega \sim \Delta $ and reflect the fact that an overall ''level
rigidity'' develops on the scale of $\Delta $.

Indeed, by definition of $\delta \rho \left( \varepsilon \right) $, 
\begin{equation}
\int \delta \rho \left( \varepsilon \right) d\varepsilon =0
\label{level-rigidity}
\end{equation}
and thus 
\begin{equation}
\int K\left( \varepsilon _{1},\varepsilon _{2}\right) d\varepsilon _{1}=\int
K\left( \varepsilon _{1},\varepsilon _{2}\right) d\varepsilon _{2}=0
\label{Int_K-epsilon}
\end{equation}
where integration is over the entire energy spectrum. The integral can be
converted to the form (\ref{Int_K}) when, given a position in the spectrum $%
\varepsilon $, the scale on which the level rigidity (\ref{level-rigidity})
develops is much smaller than $\varepsilon $. Eqs. (\ref{K_Gauss})-(\ref
{Int_K}) show that Gaussian ensembles become rigid on the scale of $\Delta $%
. This fact can be also observed from the behavior of $\Sigma $ 
\begin{eqnarray}
\Sigma \left( \varepsilon ,E\right) &=&\frac{E}{\Delta }\text{, }E\lesssim
\Delta  \label{Sigma_Gauss-1} \\
\Sigma \left( \varepsilon ,E\right) &=&C\ln \left( \frac{E}{\Delta }\right) 
\text{, }E\gg \Delta  \label{Sigma_Gauss-2}
\end{eqnarray}
where the constant $C$ depends on the specifics of a Gaussian ensemble\cite
{BFFMPW}. The behavior of $\Delta _{3}$, 
\begin{eqnarray}
\Delta _{3}\left( \varepsilon ,E\right) &=&\frac{1}{15}\frac{E}{\Delta }%
\text{, }E\lesssim \Delta  \label{Delta3_Gauss-1} \\
\Delta _{3}\left( \varepsilon ,E\right) &=&\frac{C}{2}\ln \left( \frac{E}{%
\Delta }\right) \text{, }E\gg \Delta  \label{Delta3_Gauss-2}
\end{eqnarray}
is then easily recovered from that of $\Sigma $ using (\ref{Delta3_vs_Sigma}%
). Notice that the linear terms in eqs. (\ref{Sigma_Gauss-1}) and (\ref
{Delta3_Gauss-1}) originate in the $\delta $-function term in (\ref{K_Gauss}%
) (uncorrelated levels).

In the semiclassical approach, $\delta \rho \left( \varepsilon \right) $ is
an oscillatory term that can be written as a sum over all periodic orbits%
\cite{G}. Berry\cite{B} used this approach to obtain the following limiting
behaviors of $\Delta _{3}\left( \varepsilon ,E\right) $ for the rectangular
box: 
\begin{eqnarray}
\Delta _{3}\left( \varepsilon ,E\right) &=&\frac{1}{15}\frac{E}{\Delta }%
\text{, }E\lesssim E_{\max }=\left( \pi \alpha ^{1/2}\varepsilon \Delta
\right) ^{1/2}  \label{Delta3_rectangle-1} \\
\Delta _{3}\left( \varepsilon ,E\right) &=&\frac{1}{\pi ^{5/2}}\left( \frac{%
\varepsilon }{\Delta }\right) ^{1/2}\sum_{M_{1}=0}^{\infty
}\sum_{M_{2}=0}^{\infty }\frac{\delta _{M}}{\left( M_{1}^{2}\alpha
^{1/2}+M_{2}^{2}\alpha ^{-1/2}\right) ^{3/2}}\text{, }E\gg E_{\max }
\label{Delta3_rectangle-2} \\
&\rightarrow &0.0947\sqrt{\frac{\varepsilon }{\Delta }}\text{, }\alpha \sim 1
\label{Delta3_rectangle-3}
\end{eqnarray}
where the second term gives the ''saturation rigidity'' $\Delta _{3}^{\infty
}$ and 
\begin{equation}
\delta _{M}= 
\begin{array}{l}
0 \\ 
1/4 \\ 
1
\end{array}
\begin{array}{l}
\text{if }M_{1}=M_{2}=0 \\ 
\text{if one of }M_{1}\text{ and }M_{2}\text{ is zero} \\ 
\text{otherwise}
\end{array}
\label{deltaM}
\end{equation}
Here $M_{1}$ and $M_{2}$ are the ''winding numbers'' of the classical
periodic orbits\cite{B} and the factor $\delta _{M}$ differentiates between
the self-retracing and non-self-retracing orbits. Notice that both $\Delta
_{3}^{\infty }$ and $E_{\max }$ are the functions of the position $%
\varepsilon $ in the spectrum ($\propto \sqrt{\varepsilon }$), which is in
stark contrast with Gaussian ensembles. The quantum scale $E_{\max }$ for
the onset of saturation corresponds to the time of traversal of the shortest
classical periodic orbit \cite{B}, 
\begin{equation}
E_{\max }=\frac{h}{T_{\min }}  \label{Emax}
\end{equation}
whose length is just twice the length of the rectangle's smaller side and
the winding numbers are $M_{1}=0$, $M_{2}=1$.

Notice that saturation of $\Delta _{3}$ to $\Delta _{3}^{\infty }$ in a
rectangle, eqs.(\ref{Delta3_rectangle-1}) and (\ref{Delta3_rectangle-2}), is
analogous to ''saturation'' of $\Delta _{3}$ to the weak logarithmic
dependence for a Gaussian ensemble, eqs. (\ref{Delta3_Gauss-1}) and (\ref
{Delta3_Gauss-2}). Consequently, one should expect the level density
correlation function for a rectangular box to have the form similar to (\ref
{K_Gauss}), except that the scale of ${\cal K}$ is set by $\sim \sqrt{%
\varepsilon \Delta }$. Indeed, in our previous work\cite{SW} we introduced a
simple ansatz that resulted in the following expression for $K$: 
\begin{equation}
K\left( \varepsilon ,\omega \right) =\Delta ^{-2}\left[ \delta \left( \frac{%
\omega }{\Delta }\right) -\frac{\Delta }{\pi \omega }\sin \left( \frac{2\pi
\omega }{E_{\max }}\right) \right]  \label{K_rectangle-ansatz}
\end{equation}
In this form, $K\left( \varepsilon ,\omega \right) $ satisfies eq. (\ref
{Int_K}) and is, thus, consistent with the level rigidity developing on the
scale of $E_{\max }$. Evaluation of $\Delta _{3}$ using eq. (\ref
{K_rectangle-ansatz}) successfully reproduces saturation to $\Delta
_{3}^{\infty }$ on this scale also. For $\Sigma $, eq. (\ref
{K_rectangle-ansatz}) predicts\cite{SW} eventual saturation to $\Sigma
^{\infty }=2\Delta _{3}^{\infty }$, but in an oscillatory fashion 
\begin{equation}
\frac{\Sigma -\Sigma ^{\infty }}{\Sigma ^{\infty }}=\frac{\sin \left(
E/E_{\max }\right) }{E/E_{\max }}  \label{Sigma_vs_Sigmainf}
\end{equation}
(The oscillatory term does not contribute to $\Delta _{3}^{\infty }$ in eq. (%
\ref{Delta3_vs_Sigma})).

While successfully describing the onset of the level rigidity and the
oscillations of $\Sigma $ on the scales $\sim E_{\max }$, the ansatz (\ref
{K_rectangle-ansatz}) does not capture the correct behavior of $\Sigma $ on
scales $>E_{\max }$. The central result of this work is that the magnitude
of the oscillations around $\Sigma ^{\infty }$ is not decaying and that the
onset of the level rigidity on the scale of $E_{\max }$ is not, in fact,
precise. This will be shown both numerically and analytically, the latter
being based on a more careful analysis of the level correlation function
obtained, alternatively, either semiclassically or by the direct use of the
explicit form of the level spectrum in the rectangular box.

\section{Correlation function of the level density}

The analytical expression for the level density correlation function, in the
form of the infinite sum, can be obtained in two ways. First, it can be
recovered from the semiclassical derivation of Berry in Ref.\cite{B}, as per
eqs. (23), (38), (58), and (60) there. Indeed, taking the inverse Fourier
transform in the last of these equations, we find 
\begin{eqnarray}
K\left( \varepsilon ,\omega \right) &=&\frac{1}{\sqrt{\left( \pi \Delta
\right) ^{3}\varepsilon }}\sum_{M_{1}=0}^{\infty }\sum_{M_{2}=0}^{\infty
}4\delta _{M}\frac{\cos \left( \sqrt{\frac{4\pi }{\varepsilon \Delta }\left(
M_{1}^{2}\alpha ^{1/2}+M_{2}^{2}\alpha ^{-1/2}\right) }\omega \right) }{%
\sqrt{M_{1}^{2}\alpha ^{1/2}+M_{2}^{2}\alpha ^{-1/2}}}  \label{K_rectangle}
\\
&\rightarrow &\frac{1}{\sqrt{\left( \pi \Delta \right) ^{3}\varepsilon }}%
\left[ 4\sum_{M_{1}>0}^{\infty }\sum_{M_{2}>0}^{\infty }\frac{\cos \left( 
\sqrt{\frac{4\pi }{\varepsilon \Delta }\left( M_{1}^{2}+M_{2}^{2}\right) }%
\omega \right) }{\sqrt{M_{1}^{2}+M_{2}^{2}}}+2\sum_{M>0}^{\infty }\frac{\cos
\left( \sqrt{\frac{4\pi }{\varepsilon \Delta }}M\omega \right) }{M}\right]
_{\alpha \sim 1}  \label{K_rect-apha_1}
\end{eqnarray}
The paragraph following eq. (38) in\cite{B} explains the nature of the
factor $4\delta _{M}$ in the above equation; our interpretation is that the
factor of $4$ is the intensity factor due to constructive interference
between the closed orbit and its time-reversed for non-self-retracing orbits
(both $M_{1}$ and $M_{2}$ are non-zero). The latter is crucial for
understanding of the change in level correlations due to time-reversal
symmetry breaking (for instance, by a magnetic field).

Alternatively, one can obtain (\ref{K_rectangle}) starting directly from
eqs. (\ref{rho}) and (\ref{enm}). In Ref.\cite{vO}, such a formalism was
developed for $\delta \rho $ and was generalized to $K\left( \varepsilon
,\omega \right) $ in Ref.\cite{SW}. It is based on the use of the Poisson
summation formula, whose net effect is to convert the sum over actual levels
to the sum over classical periodic orbits. Formally, it is accomplished\cite
{vO},\cite{SW} by converting the sum on $\left( n,m\right) $ to the sum on $%
\left( M_{1},M_{2}\right) $ of the Fourier transforms and by neglecting the
rapidly oscillating terms, that is harmonic terms whose arguments depend on $%
\varepsilon $ and change by $\sim \sqrt{\varepsilon /\Delta }\gg 1$ when $%
M_{1,2}$ change by $1$. As a result one recovers\footnote{%
Notice that the correct power in the denominator of eq. (20) in Ref.\cite{SW}
is $1,$ not $1/2$. Also, we are presently investigating if the coefficient $%
\delta _{M}$ is properly recovered with this technique.}\cite{SW} eq. (\ref
{K_rectangle}).

Notice that in the limit of $\omega \ll \sqrt{\varepsilon \Delta }$, the
summation over $\left( M_{1},M_{2}\right) $ can be replaced by integration
over $\left( x,y\right) $ in the entire plane, where 
\begin{equation}
x=M_{1}\alpha ^{1/4}\sqrt{\frac{4\pi \Delta }{\varepsilon }}\text{, }%
y=M_{2}\alpha ^{-1/4}\sqrt{\frac{4\pi \Delta }{\varepsilon }}
\label{xy_variables}
\end{equation}
Converting to polar coordinates, we find 
\begin{equation}
K\left( \varepsilon ,\omega \right) \rightarrow \frac{1}{2\pi ^{2}\Delta ^{2}%
}\int_{0}^{2\pi }d\phi \int_{0}^{\infty }d\rho \cos \left( \frac{\omega }{%
\Delta }\rho \right) =\Delta ^{-2}\delta \left( \frac{\omega }{\Delta }%
\right)  \label{K_rectangle-limit}
\end{equation}
that is the first term in eq. (\ref{K_rectangle-ansatz}). Notice, however,
that in reality the lower limit of $\rho $-integration should not extend to
zero since $M_{1,2}$ are not simultaneously zero. The second (repulsion)
term in eq. (\ref{K_rectangle-ansatz}) is then obtained using the ansatz\cite
{SW} in which the lower cut-off in $\int_{\rho _{\min }}^{\infty }$ is
chosen at $\rho _{\min }=y_{\min }=\alpha ^{-1/4}\sqrt{4\pi \Delta
/\varepsilon }$.

In the opposite limit $\omega \ll \sqrt{\varepsilon \Delta }$, the terms in
the series (\ref{K_rectangle}) become rapidly oscillating with the amplitude
rapidly decreasing with the increase of the winding numbers. Consequently,
the sum will be dominated by a just few terms with the smallest $M_{1,2}$.
The latter is the reason behind the non-decaying oscillations of $\Sigma $
that will be discussed below.

\section{Level number variance}

Combining eqs. (\ref{Sigma_integral}) and (\ref{K_rectangle}), we find 
\begin{eqnarray}
\Sigma \left( \varepsilon ,E\right) &=&\sqrt{\frac{\varepsilon }{\pi
^{5}\Delta }}\sum_{M_{1}=0}^{\infty }\sum_{M_{2}=0}^{\infty }4\delta _{M}%
\frac{\sin ^{2}\left( E\sqrt{\frac{\pi }{\varepsilon \Delta }\left(
M_{1}^{2}\alpha ^{1/2}+M_{2}^{2}\alpha ^{-1/2}\right) }\right) }{\left(
M_{1}^{2}\alpha ^{1/2}+M_{2}^{2}\alpha ^{-1/2}\right) ^{3/2}}
\label{Sigma_rectangle} \\
&\rightarrow &\sqrt{\frac{\varepsilon }{\pi ^{5}\Delta }}\left[
4\sum_{M_{1}>0}^{\infty }\sum_{M_{2}>0}^{\infty }\frac{\sin ^{2}\left( E%
\sqrt{\frac{\pi }{\varepsilon \Delta }\left( M_{1}^{2}+M_{2}^{2}\right) }%
\right) }{\left( M_{1}^{2}+M_{2}^{2}\right) ^{3/2}}+2\sum_{M>0}^{\infty }%
\frac{\sin ^{2}\left( E\sqrt{\frac{\pi }{\varepsilon \Delta }}M\right) }{%
M^{3}}\right] _{\alpha \sim 1}  \label{Sigma_rect-alpha_1}
\end{eqnarray}
For narrow intervals, eq. (\ref{Sigma_rectangle}) reduces to 
\begin{equation}
\Sigma \left( \varepsilon ,E\right) =\frac{E}{\Delta }\text{, }E\ll E_{\max }
\label{Sigma_rectangle-1}
\end{equation}
which follows either from eqs. (\ref{Sigma_integral}) and (\ref
{K_rectangle-limit}) or directly from (\ref{Sigma_rectangle}) if summation
is replaced with integration using variables (\ref{xy_variables}). It is
also consistent with eqs. (\ref{Delta3_vs_Sigma}) and (\ref
{Delta3_rectangle-1}).

Fig. 2 shows the result of evaluation of $\Sigma $ using eq. (\ref
{Sigma_rectangle}) (with the upper limit of summation on $M_{1,2}$ limited
to $100$) versus the numerical evaluation of $\Sigma $ for the ensemble of $%
200$ algebraic $\alpha \in \left[ 1,2\right] $. We also show $\Sigma $
obtained using the ansatz (\ref{K_rectangle-ansatz}) of Ref.\cite{SW}$.$
Clearly, the latter adequately describes the transition from the small-scale 
$E\ll E_{\max }$ linear behavior (\ref{Sigma_rectangle-1}) to scales of $%
E\sim E_{\max }$, but fails to describe the large-scale behavior,\thinspace $%
E\gg E_{\max }$. On the other hand, the agreement of numerical evaluation
with the theoretical result (\ref{Sigma_rectangle}) for all $E$ is quite
remarkable.

Two comments are in order. First, it was necessary to have a sufficient
spread of $\alpha $-values (between $1$ and $2$ here) to obtain reliable
statistics for large quantum numbers with relatively small $\alpha $%
-ensembles ($200$ values here). However, the dependence on $\alpha $ of the
relevant parameters is quite weak, through $\alpha ^{\pm 1/4}$, so for $%
\alpha \in \left[ 1,2\right] $ the difference from an ensemble whose $\alpha 
$-values would be close to a fixed value ($1$ in this particular case) is
not significant. Second, for $E\gg E_{\max }$ it is sufficient to limit
summation on $M_{1,2}$ to $2$ and $3$ in the double and single sum
respectively to obtain a curve which is very close to the full theoretical
curve; as was already explained above this is because the amplitudes of the
quickly oscillating terms in the sum rapidly fall off with the increase of $%
M_{1,2}$.

We now turn to the theoretical curve in Fig. 3 that represents the level
number variance given by (\ref{Sigma_rect-alpha_1}). It is intended to
describe the $\alpha $-ensemble such that $\alpha \simeq 1$ and it can be
effectively reproduced by the superposition of only $6$ harmonics: $3$ from
the double some and $3\ ($with commensurate frequencies) from the single
sum. These are non-decaying oscillations whose amplitude is given by 
\begin{equation}
\frac{\Sigma ^{\pm }}{\Sigma ^{\infty }}\approx \frac{2}{3}
\label{SigmaA_vs_Sigmainf}
\end{equation}
where $\Sigma ^{+}=\max \left\{ \Sigma \right\} -\Sigma ^{\infty }$, $\Sigma
^{-}=\Sigma ^{\infty }-\min \left\{ \Sigma \right\} $ and $\Sigma ^{\infty }$
is the mean value of oscillating $\Sigma $ which is obtained by substituting 
$\sin ^{2}\rightarrow 1/2$ in eq. (\ref{Sigma_rectangle}). The small
asymmetry of the positive and negative swings of $\Sigma $-oscillations, $%
\Sigma ^{+}\neq \Sigma ^{-}$, is due to the asymmetry of the Clausen function%
\cite{LW} $Cl_{3}(x)$, 
\[
Cl_{3}(x)=\sum_{k=1}^{\infty }\frac{\cos kx}{k^{3}} 
\]
for which $\max \left\{ Cl_{3}(x)\right\} /\min \left\{ Cl_{3}(x)\right\}
\approx 1.3$. This function originates in the single sum in eq. (\ref
{Sigma_rect-alpha_1}), which corresponds to self-retracing orbits.

It is important to point out that if one averages over the range of $\alpha $%
-values, or over the range of $\varepsilon $-values\cite{Gr}, the beats that
result from a superposition of a continuous range of harmonics will present
themselves, over relevant interval widths, as decaying oscillations. This is
due to the corresponding continuous range of the beat frequencies
represented by their envelopes.

We now turn to breaking of time reversal symmetry for charged particles due
to a magnetic field. Clearly, the condition for the latter is given by 
\begin{equation}
BA\sim \phi _{0}  \label{Tbreak}
\end{equation}
where $\phi _{0}=hc/e$ is the flux quantum. For such fields, the Larmor
radius is much greater than either side of the rectangle 
\begin{equation}
R=\frac{mcv}{eB}\sim \sqrt{A}\sqrt{\frac{\varepsilon }{\Delta }}
\label{Larmor}
\end{equation}
where $mv^{2}/2=\varepsilon $. Therefore, the deviation from the free
specular scattering will be small and we can find both the level correlation
function and the level number variance via simple substitution 
\begin{equation}
\delta _{M}\rightarrow \delta _{M}^{\prime }= 
\begin{array}{l}
0 \\ 
1/4 \\ 
1/2
\end{array}
\begin{array}{l}
\text{if }M_{1}=M_{2}=0 \\ 
\text{if one of }M_{1}\text{ and }M_{2}\text{ is zero} \\ 
\text{otherwise}
\end{array}
\label{deltaMp}
\end{equation}
in eqs. (\ref{K_rectangle}) and (\ref{Sigma_rectangle}). For the $\alpha
\simeq 1$-ensemble considered above, $\Sigma ^{\prime }$ is shown in Fig. 4.
In this case 
\begin{eqnarray}
\frac{\Sigma ^{\prime \infty }}{\Sigma ^{\infty }} &\approx &0.7
\label{Sigmainfp_vs_Sigmainf} \\
\frac{\Sigma ^{\prime +}}{\Sigma ^{\prime \infty }} &\approx &0.77\text{, }%
\frac{\Sigma ^{\prime -}}{\Sigma ^{\prime \infty }}\approx 0.64
\label{SigmainfAp_vs_Sigmainfp}
\end{eqnarray}
The increased asymmetry of $\Sigma $-oscillations underscores the increased
relative contribution of the single sum to $\Sigma $ in eq. (\ref
{Sigma_rect-alpha_1}) (self-retracing orbits).

\section{Discussion}

The central results of this work are summarized by Figs. 2-4. The first of
these graphs shows that the numerical evaluation of the level number
variance for a particle in a rectangular box is in excellent agreement with
the theoretical result given by eq. (\ref{Sigma_rectangle}). The second
indicates a non-decaying oscillatory behavior of $\Sigma $. The third shows $%
\Sigma $ when the time-reversal symmetry is broken. Figs. 3 and 4 can be
successfully reproduced with just a small number of lowest harmonics in (\ref
{Sigma_rect-alpha_1}).

It is remarkable that, given the center of the interval $\varepsilon $, $%
\Sigma \left( \varepsilon ,E\right) $ exhibits large, reproducible
oscillations as a function of the interval width $E$. The main implication
of this result is that while the level rigidity indeed develops on the scale
set by $\sim \sqrt{\varepsilon \Delta }\ll \varepsilon $, it is accurate
only in an approximation where a harmonic is replaced by its average - zero.
In other words, (\ref{Int_K}) converges to zero on the scale of $\omega \sim 
\sqrt{\varepsilon \Delta }$ only up to a sum of harmonic terms, as implied
by eq. (\ref{K_rectangle}).

We point out that $\Sigma $-oscillations are entirely consistent with the
near straight line saturation of $\Delta _{3}$. In fact, $\Delta _{3}$ also
exhibit oscillatory behavior around $\Delta _{3}^{\infty }$, but it is both
parametrically small in $\Delta /\varepsilon $ and decays as a function of $E
$. This reduction of $\Delta _{3}$ can be traced, for instance, to
integration in eq. (\ref{Delta3_vs_Sigma}).

In the future, we will investigate the level number variance in a variety of
finite-domain and potential problems. We will also address the explicit
dependence of the level correlation function on the magnetic field for a
detailed description of the time-reversal breaking transition and to address
the orbital magnetism of non-resonant integrable systems.

\section{Acknowledgments}

We are grateful to Bernie Goodman for numerous discussions and insightful
comments.

\section{Figure captions}

\subsubsection{Figure 1}

Level rigidity $\Delta _{3}\left( \varepsilon ,E\right) $ vs. the interval
width $E$ for $\varepsilon /\Delta =10^{4}$ and $200$ algebraic $\alpha \in
\left[ 1,2\right] $. The straight line is $\left( E/15\Delta \right) $ and
the two horizontal lines correspond to Berry's saturation rigidity $\Delta
_{3}^{\infty }$ for $\alpha =1$ (lower line) and $\alpha =2$. Notice that
our saturation rigidity is slightly below the former, while expected to be
between the two, and that the difference from $\Delta _{3}^{\infty }$ is
amplified by the factor of $\sqrt{\varepsilon /\Delta }$.

\subsubsection{Figure 2}

Numerical evaluation of $\Sigma \left( \varepsilon ,E\right) $ (red line),
theoretical evaluation per eq. (\ref{Sigma_rectangle}) (green line), and
theoretical evaluation based on ansatz (\ref{K_rectangle-ansatz}) (blue
line) for $\varepsilon /\Delta =10^{4}$ and $200$ algebraic $\alpha \in
\left[ 1,2\right] $.

\subsubsection{Figure 3}

$\Sigma \left( \varepsilon ,E\right) $ per eq. (\ref{Sigma_rect-alpha_1})
for $\varepsilon /\Delta =10^{5}$.

\subsubsection{Figure 4}

$\Sigma \left( \varepsilon ,E\right) $ per eqs. (\ref{Sigma_rect-alpha_1})
and (\ref{deltaMp}) for $\varepsilon /\Delta =10^{5}$ for the case of broken
time reversal symmetry.

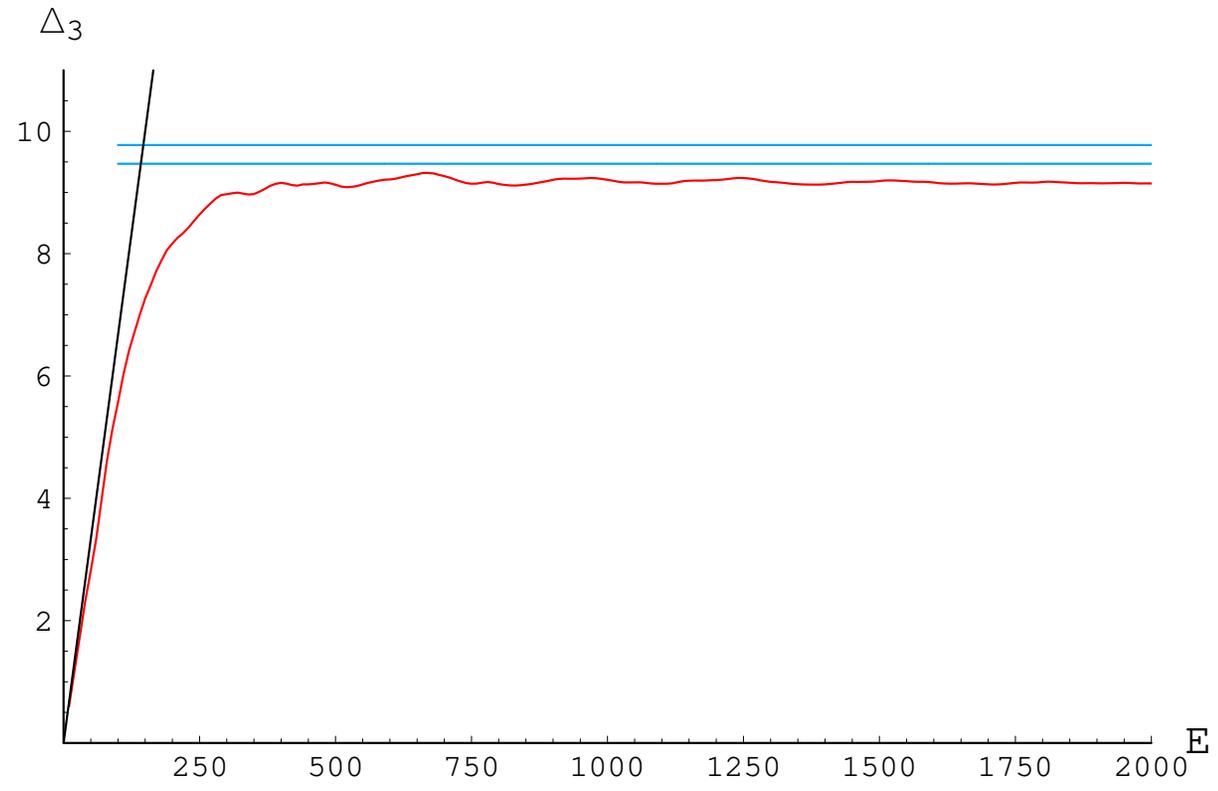
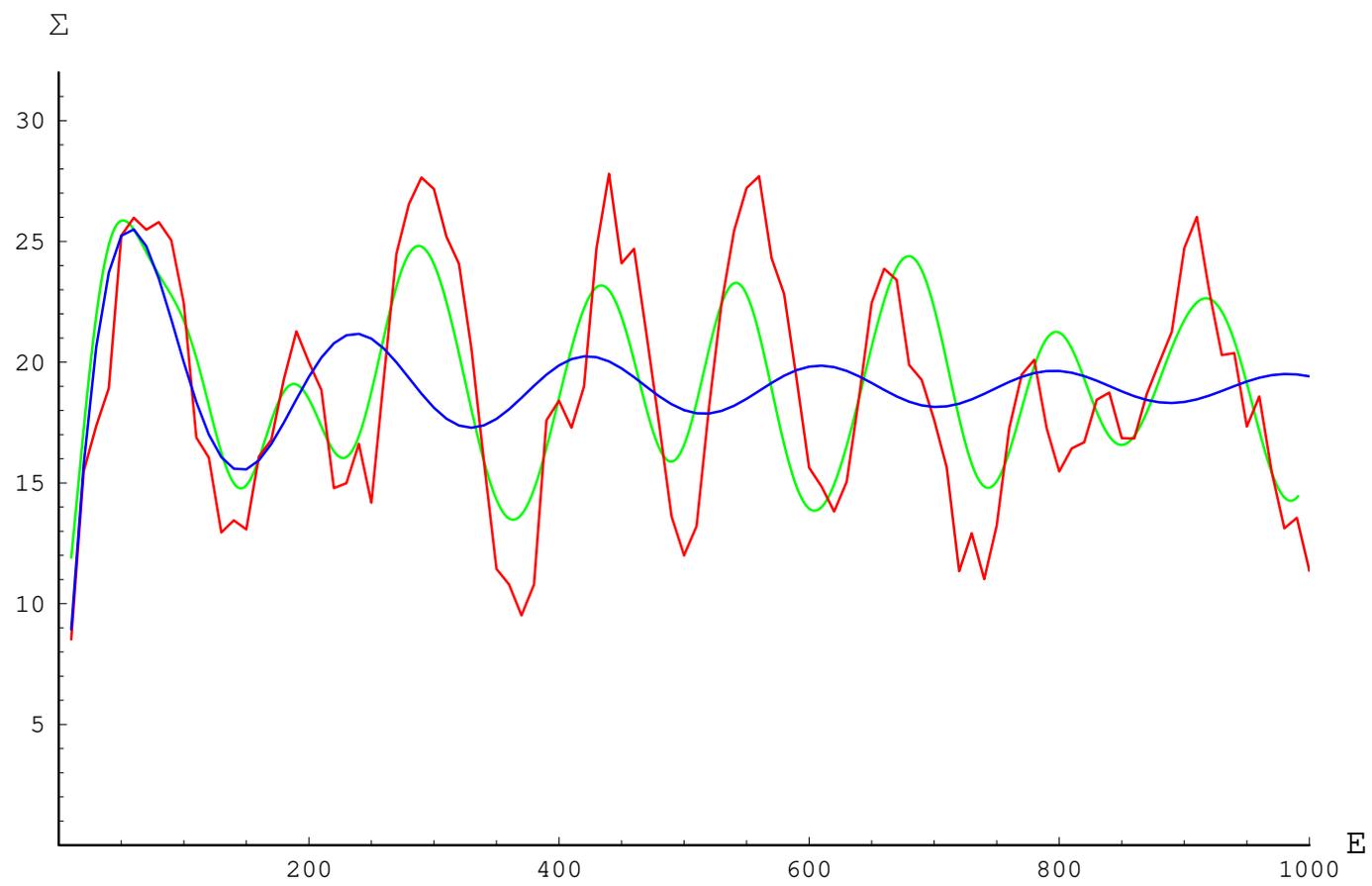
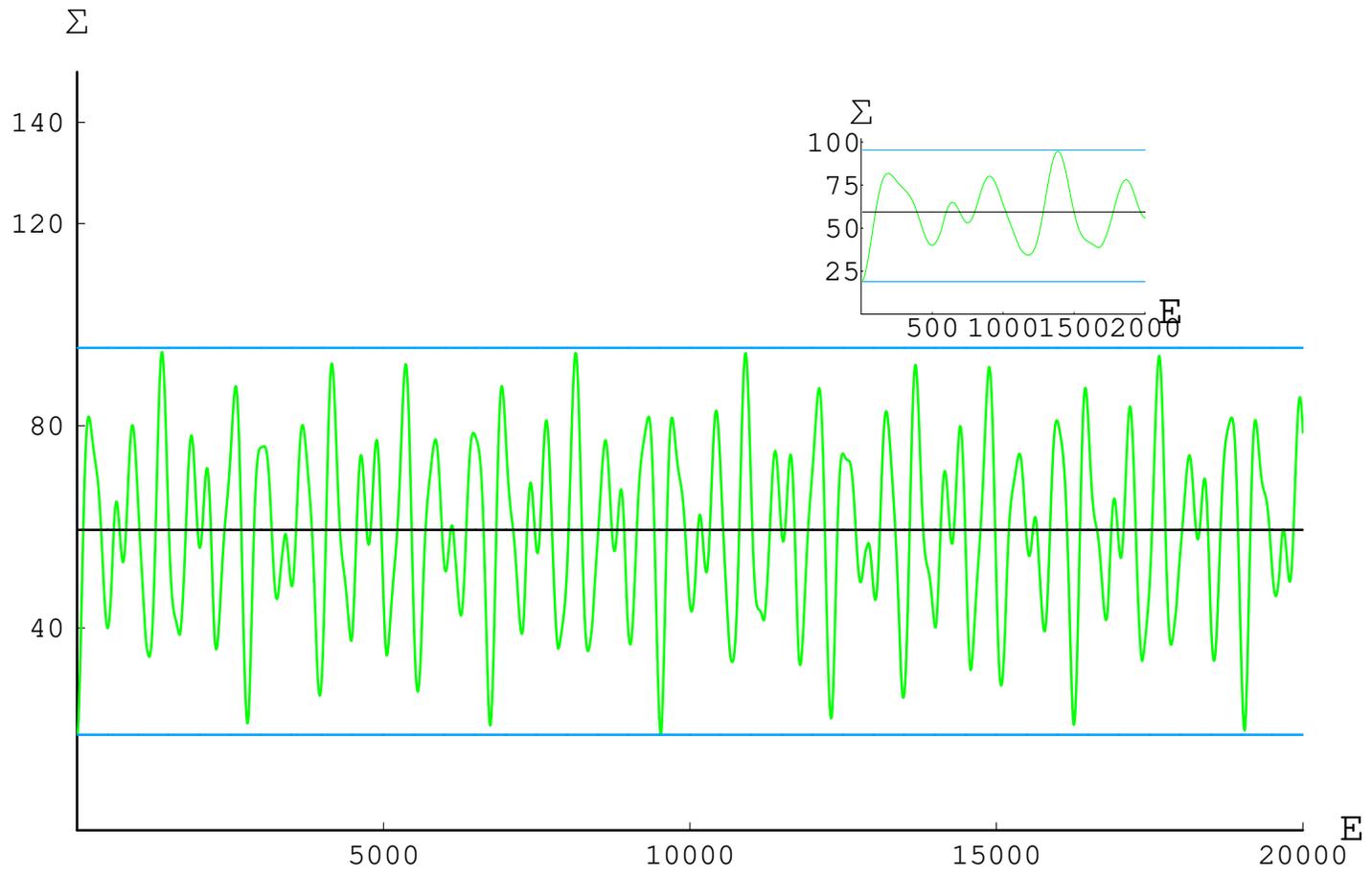
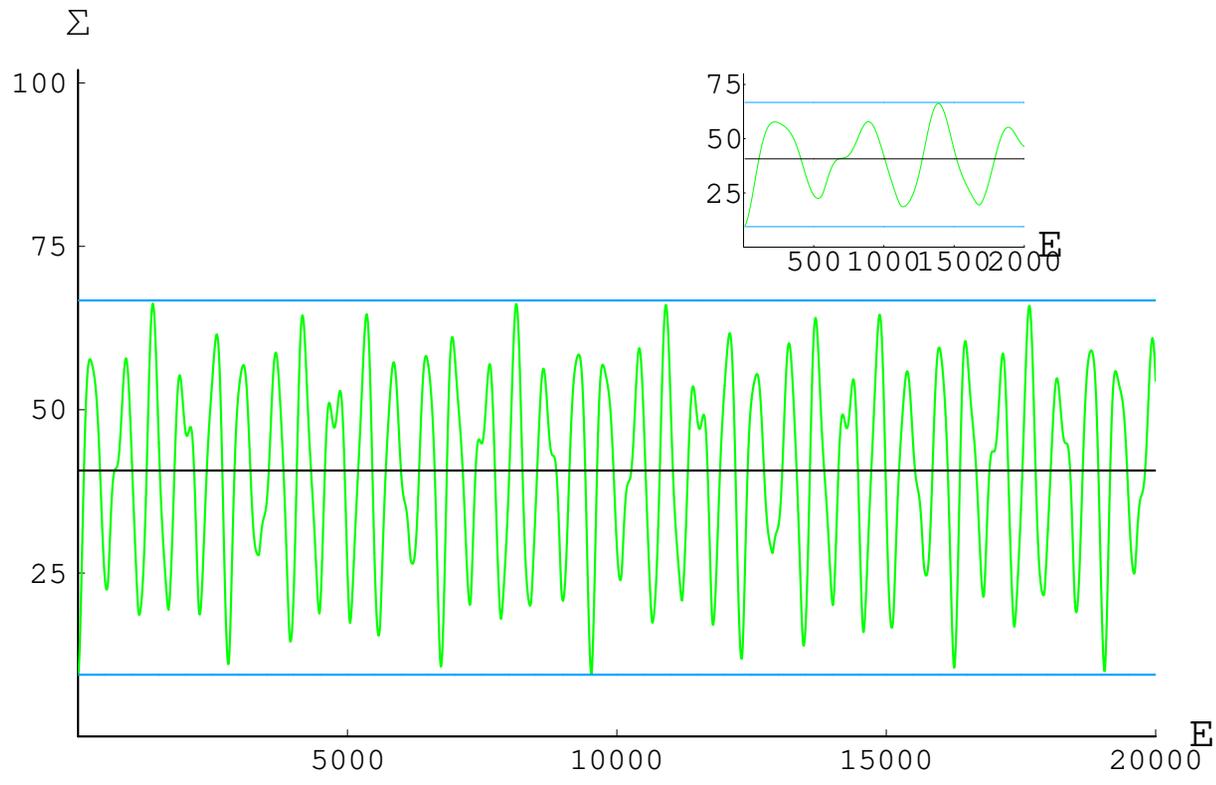
\end{document}